\documentclass[12pt, preprint]{aastex}

\slugcomment{Astrophysical Journal, August 1, \textbf{664},
pp.??-??}

\shorttitle{Over-reflection and generation of Gravito-Alfv\'en
waves}

\shortauthors{Rogava et al.}

\begin{document}

\title{On over-reflection and generation of Gravito-Alfv\'en waves
in solar-type stars}

\author{Andria Rogava\altaffilmark{1}}
\affil{Centre for Plasma Astrophysics, K.U.Leuven, Celestijnenlaan
200B, 3001 Leuven, Belgium; Abdus Salam International Centre for
Theoretical Physics, Trieste I-34014, Italy.}
\email{Andria.Rogava@wis.kuleuven.be}

\author{Grigol Gogoberidze\altaffilmark{1}}
\affil{Centre for Plasma Astrophysics, K.U.Leuven, Celestijnenlaan
200B, 3001 Leuven, Belgium.}
\email{Grigol.Gogoberidze@wis.kuleuven.be }

\author{Stefaan Poedts}
\affil{Centre for Plasma Astrophysics, K.U.Leuven, Celestijnenlaan
200B, 3001 Leuven, Belgium.}
\email{Stefaan.Poedts@wis.kuleuven.be}

\altaffiltext{1}{On leave from the Georgian National Astrophysical
Observatory, Kazbegi ave. $2^a$ Tbilisi-0160, Georgia.}

\begin{abstract}
The dynamics of linear perturbations is studied in magnetized
plasma shear flows with a constant shearing rate and with
gravity-induced stratification. The general set of linearized
equations is derived and the two-dimensional case is considered in
detail. The \emph{Boussinesq approximation} is used in order to
examine relatively small-scale perturbations of low-frequency
modes: \emph{Gravito-Alfv\'en waves} (GAW) and \emph{Entropy Mode}
(EM) perturbations. It is shown that for flows with arbitrary
shearing rate there exists a finite time interval of
\emph{non-adiabatic} evolution of the perturbations. The
non-adiabatic behavior manifests itself in a twofold way, viz.\ by
the over-reflection of the GAWs and by the generation of GAWs from
EM perturbations. It is shown that these phenomena act as
efficient transformers of the equilibrium flow energy into the
energy of the perturbations for moderate and high shearing rate
solar plasma flows. Efficient generation of GAW by EM takes place
for shearing rates about an order of magnitude smaller than
necessary for development of a shear instability. The latter fact
could have important consequences for the problem of angular
momentum redistribution within the Sun and solar-type stars.
\end{abstract}


\section{Introduction}

It is well-known that the most advanced stellar models, taking
into account rotation-induced hydrodynamic processes of meridional
circulation and shear mixing, are quite adequately describing the
structure of massive stars \citep{mm00}. For solar-type stars with
relatively wide convective regions, however, these models predict
\emph{large} rotation gradients and are in a notable contradiction
with recent observational helioseismology data \citep{ct05}. This
is probably due to the insufficiently high efficiency of the
common hydrodynamic instabilities that are characteristic to these
models: these instabilities can redistribute the angular momentum
between the different layers of the radiative region, but they are
\emph{not} strong enough to ensure the onset of a uniform rotation
regime.

While the radiative interiors of solar-type stars rotate quite
uniformly, their convective zones rotate differentially. Mounting
observational evidence indicates that the radiative interior of
the Sun, for instance, rotates almost as a solid body, with a
quasi-flat seismic rotation profile \citep{c99,c03}. But the solar
convective zone exhibits a strong shearing in the latitudinal
direction, with its equatorial layers rotating about 1/3 faster
than the polar regions \citep{km03}. The transition between the
differential and the uniform rotation regimes takes place in a
relatively thin layer (with thickness $\le 0.05 R_{\bigodot}$),
called the \emph{tachocline}, where the radial shearing of the
rotation is particularly significant \citep{m05,p03}. It is often
argued that some kind of shear instability must be the source of
gravity waves at the base of the solar convection zone
\citep{k99}, either in the immediate vicinity or just inside the
tachocline. This instability can hardly be of a purely
hydrodynamical nature, because from various numerical studies of
the buoyant rise of magnetic flux from the bottom of the
convection zone to the surface of the sun it follows that the
tachocline is strongly magnetized with its poloidal magnetic field
about $10^4-10^5\;$G \citep{m05}.

From this perspective it seems quite plausible to surmise that on
the time-scale of the order of the solar age or less not only
purely hydrodynamic internal gravity waves (IGWs) participate in
the process of the angular momentum redistribution within the
solar interior, but also, or rather, Gravito-MHD waves. This idea
is \emph{not} new. In the literature it has been argued
\citep{z97,kq97} that Gravito-MHD waves could be one of the most
promising candidates for the onset of the redistribution
processes. It was argued that a self-consistent model should
comprise a large-scale magnetic field in the Sun's interior, and
an accurate consideration of the Coriolis effects in the
convection zone and in the tachocline
\citep{tc98,tc03,tc04,tc05,ct05}. It was also argued that
turbulent stresses in the convection zone, via the agency of
Coriolis effects, induce a meridional circulation, causing the gas
from the convection zone to burrow downwards and generate both
horizontal and vertical velocity shear characterizing the
tachocline. However, the interior magnetic field confines the
burrowing and, hence, smooths and diminishes the shear, as
demanded by the observed structure of the tachocline. According to
\citet{ct05}, a decisive test of this \emph{qualitative} scenario,
after numerical refinements, would be its \emph{quantitative}
consistency with the observed interior angular velocity structure.


The aim of the present paper is to present the results of a
detailed study of the linear dynamics of perturbations in
gravitationally stratified magnetized shear flows. Bearing in mind
the highly probable importance of the Gravito-MHD waves for the
structure of the Sun and solar-type stars, we decided to pose and
study the following three interrelated issues: 1)~how the presence
of the nonuniform velocity field affects the propagation of the
waves through the stellar plasma; 2)~what kind of energy exchange
processes between the different collective modes and between the
modes and the ambient flow may happen; and 3)~what other
astrophysical consequences these processes could have.

Originally, we develop the three-dimensional (3D) model, allowing
the ambient flow to have velocity shearing in both transversal
(prior to the gravity field) directions. Further on, the study is
focused by a number of simplifying assumptions, enabling the
solution of the equations. First of all, we consider only two
spatial dimensions (2D) bearing in mind that it is quite
straightforward to extend the analysis to the fully
three-dimensional case. Second, we assume a constant steady-state
(or stationary) shear flow, an assumption that allows us to use
the \emph{shearing sheet approximation} \citep{gl65}. And third,
we use the \emph{Boussinesq approximation}, and hence we study the
dynamics of relatively small-scale\footnote{with the vertical
wavelength, $\ell_z$, much smaller then stratification length
scale $z_0$.} perturbations of the low-frequency modes, viz.\ the
Gravito-Alfv\'en waves (GAW) and the Entropy Mode (EM)
perturbations.

Our study leads to the conclusion that for \emph{arbitrary} shear
in the flow, there exists a finite interval of time in which the
evolution of the modes is highly non-adiabatic. Outside this
interval, the evolution is adiabatic and can be accurately
described by a WKB approximation.  Therefore, we have an
asymptotic problem, quite common for various quantum-mechanical
applications, where one needs to obtain connection formulae in
\emph{turning point problems} \citep{ll77}. The relevant analysis
is done in this paper and it is found that the non-adiabatic
behavior of the considered modes manifests itself in the form of
two phenomena, viz.\ the \emph{over-reflection} of the GAW and the
\emph{generation} of the GAW by the EM perturbations. We show that
for flows with moderate and high shearing rates both these
processes are very effective converters of the equilibrium flow
energy into the energy of the waves.

In the context of above-discussed astrophysical problems our
results imply that efficient generation of the GAW by the EM
perturbations takes place for shearing rates of about an order of
magnitude smaller then necessary for development of shear
instability.

The remainder of the paper is organized in the following way: the
general mathematical and heuristic formalism is presented in the
Sec.~2. The over-reflection of GAW is studied in the Sec.~3; and
the second non-adiabatic process, viz.\ the generation of the GAW
by the EM perturbations, is studied in the Sec.~4. Finally, in
Sec.~5 we conclude with a summary and a brief discussion of our
results in the context of their possible importance for the
angular momentum redistribution within solar-type stars and the
solution of the problem of the uniform rotation of their radiative
zones.

\section{Linear perturbations of sheared plasma flows}

For the sake of generality, we derive the basic set of equations
for the case of 3D perturbations in a compressible,
gravitationally stratified MHD fluid. This formalism will be
presented in the next (2.1) subsection. Later on (subsection 2.2),
we will restrict the consideration to the somewhat simpler case of
2D perturbations in an incompressible medium. Besides, the
\emph{Boussinesq approximation} will be used and the stationary
shear flow will be considered. 

\subsection{General theory}

In our model, the geometry of the considered problem is simplified
in the following way: the equilibrium flow ${\bf {U}}$ is supposed
to be plane-parallel, to be directed along the $x$-axis, and to
have both a horizontal ($A_y$) and a vertical ($A_z$) shear:
$$
{\bf {U}} \equiv \left[ A_y y + A_z z, 0, 0  \right]. \eqno(1)
$$

The uniform gravity ${\bf g}$ is assumed to be directed along the
negative direction of the $z$-axis:
$$
{\bf g} \equiv \left[ 0, 0, -g \right]. \eqno(2)
$$

We consider a simplified model and assume that the equilibrium
magnetic field ${\bf B_0}$ is toroidal, parallel to $\bf {U}$ and
that it is possessing the gravity-induced vertical stratification:
$$
{\bf B_0} \equiv \left[ B(z),~0,~0~ \right]. \eqno(3)
$$

The set of equations of one-fluid ideal magnetohydrodynamics
(MHD), governing the physics of the flow is:
$$
D_t \rho + \rho (\nabla \cdot \bf V)=0, \eqno(4)
$$
$$
\rho D_t{\bf V}= -{\nabla}{\left(P+{{B^2}\over{8{\pi}}}\right)}
+{{1}\over{4{\pi}}}({\bf B}\cdot{\nabla}){\bf B}+ \rho {\bf g},
\eqno(5)
$$
$$
D_t \bf B= (\bf B \cdot \nabla){\bf V} - \bf B (\nabla \cdot \bf
V), \eqno(6)
$$
$$
D_t S=0, \eqno(7)
$$
while
$$
\nabla \cdot \bf B = 0. \eqno(8)
$$

In (4--7) the total (convective) time derivative operator is
denoted by: $D_t \equiv \partial_t + ({\bf V} \cdot \nabla)$.

Instantaneous values of all physical variables are decomposed into
their regular (mean) and perturbational components:
$$
\rho \equiv \rho_0 + \varrho, \eqno(9)
$$
$$
P \equiv P_0 + p, \eqno(10)
$$
$$
S \equiv S_0 + s, \eqno(11c)
$$
$$
{\bf V} \equiv {\bf U} + {\bf u}, \eqno(12a)
$$
$$
{\bf B} \equiv {\bf B}_0 + {\bf b}. \eqno(12b)
$$

It is straightforward to see that the horizontal background flow
does not affect the vertical hydromagnetic equilibrium of the
flowing MHD fluid, governed by the equation:
$$
\partial_z{\left(P_0 + {{B_0^2}\over{8\pi}} \right)} =-\rho_0g.
\eqno(13)
$$

In general the variation of the density $d \rho$ is related with
variations of the pressure $dP$ and entropy $ds$ via the relation:
$$
d \rho = \mu ds + (1/C_s^2)dp, \eqno(14)
$$
where:
$$
\mu \equiv (\partial \rho/\partial s)_p, \eqno(15)
$$
$$
C_s^{2} \equiv (\partial P/\partial \rho)_s. \eqno(16)
$$

Linearized set of equations governing the evolution of
perturbations within this flow can be written in the following way
[${\cal D}_t \equiv \partial_t + (A_y y + A_z z)\partial_x$]:
$$
{\cal D}_t \varrho + (\partial_z \rho_0) v_z + \rho_0(\nabla \cdot
{\bf u}) = 0, \eqno(17)
$$
$$
{\cal D}_t u_x = - {\cal A}u_y - Au_z
-{{1}\over{\rho_0}}\partial_xp
+{\left[{{(\partial_zB_0)}\over{4\pi}\rho_0}\right]}b_z, \eqno(18)
$$
$$
{\cal D}_t u_y = -{{1}\over{\rho_0}}\partial_yp
-{\left({{B_0}\over{4\pi
\rho_0}}\right)}(\partial_xb_y-\partial_yb_x), \eqno(19)
$$
$$
{\cal D}_t u_z = -{{1}\over{\rho_0}}\partial_zp +
{\left({{B_0}\over{4\pi}\rho_0}\right)}(\partial_xb_z-\partial_zb_x)
$$
$$
-{\left[{{(\partial_zB_0)}\over{4\pi}\rho_0}\right]}b_x
-{{g}\over{\rho_0}}\varrho, \eqno(20)
$$
$$
{\cal D}_t s + (\partial_z s) v_z = 0, \eqno(21)
$$
$$
{\cal D}_tb_y=B_0\partial_xu_y, \eqno(22)
$$
$$
{\cal D}_tb_z=B_0\partial_xu_z, \eqno(23)
$$
$$
\partial_xb_x + \partial_yb_y + \partial_zb_z = 0. \eqno(24)
$$
Due to the stratification some coefficients on the right hand
sides of Eqs. (17-23) depend on $z$. But when studying the
dynamics of small-scale perturbations (with $\ell_z \ll z_0$) one
can consider the mean components as constant. In this case the set
of first-order, partial differential equations (17-24) can be
reduced to the set of ordinary differential equations (ODEs) with
time-dependent coefficients if we look for solutions in the
following form:
$$
F({\bf r},t) \equiv {\hat F}[{\bf k}(t),t]e^{i{\left[({\bf k}
\cdot {\bf r})- \varphi({\bf k},t) \right]}}, \eqno(25)
$$
with
$$ {\varphi}({\bf k},t) \equiv {\left( {\bf U} \cdot
{\int}{\bf k}(t)dt\right)}, \eqno(26)
$$
where ${\bf k}(t)$ is a time-dependent wavenumber vector,
determined by the following set of equations \citep{lpl84, cc86,
mr99}:
$$
{\bf k}^{(1)}+{\cal S}^T{\cdot}~{\bf k}=0, \eqno(27)
$$
with ${\cal S}^T$ being the transposed \emph{shear matrix}:
$$
{\cal S}{\equiv}{\left(\matrix{U_{x,x}&U_{x,y}&U_{x,z}\cr
        U_{y,x}&U_{y,y}&U_{y,z}\cr
        U_{z,x}&U_{z,y}&U_{z,z}\cr}\right)}, \eqno(28)
$$
which, in our case has only two nonzero components: $A_y$ and
$A_z$:
$$
{\cal S}{\equiv}{\left(\matrix{0&A_y&A_z\cr
        0&0&0\cr
        0&0&0\cr}\right)}. \eqno(29)
$$
Due to Eq.~(27), transversal components of the ${\bf k}(t)$
acquire linear time-dependence:
$$
k_y(t) = k_y(0) - A_ytk_x, \eqno(30a)
$$
$$ k_z(t) = k_z(0) - A_ztk_x. \eqno(30b)
$$

Therefore, we see that the wave-number vector ${\bf k}(t)$ of the
Spatial Fourier Harmonics (SFH) varies in time, i.e.\ in the
linear approximation there is a ``drift" of the SFH in the {\bf
k}-space. The physical reason of this ``drift" is related to the
fact that in the sheared flow the perturbations cannot have the
form of a simple plane wave due to the effect of the shearing
background on the wave crests.

Applying the ansatz (25) to Eqs.~(17-23), we reduce the system to
the following set of first-order ODEs [hereafter we are using the
following notation $f^{(n)} \equiv \partial_t^{n}f$]:
$$
d^{(1)}=i \varepsilon v_z + v_x + {\cal K}_y(\tau) v_y + {\cal
K}_z(\tau) v_z, \eqno(31)
$$
$$
v_x^{(1)}= -{\cal R} v_y - R v_z - \sigma^2{\cal P} +
i{{\varepsilon}\over{2}} b_z, \eqno(32)
$$
$$
v_y^{(1)}=  - \sigma^2{\cal K}_y(\tau) {\cal P} + b_y - {\cal
K}_y(\tau) b_x, \eqno(33)
$$
$$
v_z^{(1)}= - \sigma^2{\cal K}_z(\tau) {\cal P} + b_z - {\cal
K}_z(\tau) b_x -
$$
$$
i{{\varepsilon}\over{2}} b_x + i \alpha d, \eqno(34)
$$
$$
e^{(1)} = v_z, \eqno(35)
$$
$$
b_y^{(1)} = - v_y, \eqno(36)
$$
$$
b_z^{(1)} = - v_z, \eqno(37)
$$
while the Maxwell equation (24) reduces to the following algebraic
relation:
$$
b_x + {\cal K}_y(\tau) b_y + {\cal K}_z(\tau) b_z = 0. \eqno(38)
$$

In these equations, the following dimensionless notation
\footnote{The Alfv\'en speed $C_A^2 = B_0^2/4 \pi \rho_0$ and the
Brunt-V\"ais\"al\"a frequency $N_{BV}^2=-g\partial_z
\rho_0/\rho_0$ are defined in the usual way.} is used for the
constants: $R_y \equiv A_y/k_xC_A$, $R_z \equiv A_z/k_xC_A$,
${\cal K}_y(0) \equiv k_y(0)/k_x$, ${\cal K}_z(0) \equiv
k_z(0)/k_x$, $\varepsilon \equiv (k_z z_0)^{-1}$, $\sigma^2 \equiv
(C_s/C_A)^2$, $N^2 \equiv (N_{BV}/C_Ak_x)^2$. The dimensionless
variables appearing in the above set of equations are defined as:
$v_j \equiv {\hat u}_j/C_A$, $b_j \equiv i{\hat b}_j/B_0$, $d
\equiv i{\hat \varrho}/\rho_0$, ${\cal P} \equiv i{\hat
p}/\rho_0C_s^2$, $e \equiv -k_x {\hat s}/(\partial_zs_0)$, $\tau
\equiv k_xC_At$, ${\cal K}_y(\tau) \equiv {\cal K}_y(0) - {\cal
R}\tau$, ${\cal K}_z(\tau) \equiv {\cal K}_z(0) - R\tau$.

Notice that this system is \emph{not} closed, because we have only
\emph{eight} equations for \emph{nine} variables (${\bf v}, {\bf b},
d, e, \cal P$). The closure of the set of equations is ensured by
the thermodynamic relation that follows from Eq.~(14) [$\alpha
\equiv g/k_xC_s^2$]:
$$
d = {\cal P} - i (N^2/\alpha) e. \eqno(39)
$$

This system of equations describes the temporal evolution of
Gravito-MHD waves modified by the presence of a velocity shear in
the two planes transversal to the flow direction. The full
analysis of this set of equations is beyond the scope of the
present paper. Instead, in the next sections, we focus our study
on the relatively simple 2D and incompressible case. We will see
that even in this simplified case the presence of the velocity
shear brings a considerable novelty in the dynamics of
perturbations.

\subsection{The incompressible, 2D limit}

In order to simplify the mathematical aspects of the problem, we
make the following assumptions:

\begin{itemize}
\item We consider the 2D case, restricting the problem to the
$x0z$ plan and implying:
$$
{\cal K}_y(0) = R_y =0. \eqno(40)
$$
Hereafter we write $R \equiv R_z (A \equiv A_z)$ In this case,
Eq.~(38) enables us to express the longitudinal ($x$-)component of
the magnetic field perturbation in terms of its transversal
($z$-)component:
$$
b_x = - {\cal K}_z(\tau) b_z. \eqno(41)
$$

\item We drop the $d^{(1)}$ term in the continuity equation, i.e.\ we
adopt the concept of so-called {\it dynamic incompressibility}
\citep{ll59}. This implies that the velocity of the flow is assumed
to be small enough in comparison with the speed of sound. As a
consequence, the density perturbations evoked by the pressure
perturbations are negligible and instead of the Eq.~(39) we will
have:
$$
d \simeq - i (N^2/\alpha) e. \eqno(42)
$$

However, since the medium is thermally conducting its density
should vary also due to the temperature variation, and this effect
should be (and is) taken into account.
\end{itemize}

In hydrodynamics, the second assumption is known as the
\emph{Boussinesq approximation}: in the conservation equations the
terms proportional to the density perturbation are retained if,
and only if, they produce a buoyancy force \footnote{The last term
in Eq.~(34).}. The Boussinesq approximation automatically implies
\citep{ll59,g82} that the vertical length scale of perturbations,
$\ell_z \sim k_z^{-1}$, is much smaller then the characteristic
stratification length scale $z_0$. This condition was already used
for the derivation of Eqs. (31-37). In terms of our dimensionless
notation, this condition transcribes into:
$$
\varepsilon \ll 1. \eqno(43)
$$
When this condition is satisfied, the above approach enables us to
study the dynamics of small-scale, low-frequency perturbations.
\footnote{for detailed analysis of the Boussinesq approximation in
stratified shear flows see \citet{d97} and \citet{dkp04}.}

These assumptions make the basic set of equations much simpler.
Taking into account (41) and (42), we get [with ${\cal K}^2(\tau)
\equiv 1 + {\cal K}_z^2(\tau)$]:
$$
v_x + {\cal K}_z(\tau) v_z=0, \eqno(44)
$$
$$
v_x^{(1)}= - R v_z - \sigma^2{\cal P}, \eqno(45)
$$
$$
v_z^{(1)}= - \sigma^2{\cal K}_z(\tau) {\cal P} + {\cal K}^2(\tau)
b_z + N^2 e, \eqno(46)
$$
$$
e^{(1)} = v_z, \eqno(47)
$$
$$
b_z^{(1)} = - v_z. \eqno(48)
$$

From the last two equations it is apparent that there is an
algebraic relation between the entropy perturbations and the
magnetic field perturbations, viz.\
$$
e + b_z = {\cal I}=const(\tau). \eqno(49)
$$

Another, less obvious, relation between the perturbed quantities
may be found if one takes time derivative of (44) and takes into
account (45) and (46). The result can be used to express the
pressure perturbation as a function of the other variables, i.e.\
$$
\sigma^2{\cal P}={{-2R}\over{{\cal K}^2(\tau)}}v_y+{\beta}b_y
-{{W^2{\beta}}\over{{\cal K}^2(\tau)}}e. \eqno(50)
$$

Having three algebraic relations, Eqs.~(44), (49), and (50), for
five variables, we can reduce the system to {\it one} second-order
ODE. For instance, for the variable $b(\tau)$, defined as $b(\tau)
\equiv {\cal K}(\tau)b_z$, we can derive the following explicit,
second-order inhomogeneous ODE:
$$
\frac{{\rm d^2} b}{{\rm d} \tau^2} + \left[1  + \frac{ N^2}{{\cal
K}^2(\tau)} - \frac{R^2}{{\cal K}^4(\tau)}\right] b =
\frac{N^2}{{\cal K}(\tau)} {\cal I}. \eqno(51)
$$

This equation describes the dynamics of the so-called
\emph{continuous eigenspectrum} (see, {\it e.g.} \cite{bm02} and
the references therein). In order to take into consideration the
perturbations belonging to the \emph{discrete eigenspectrum} one
should specify some boundary conditions with respect to the
vertical spatial variable $z$. In the absence of an external
magnetic field, it is known \citep{dr81} that all the discrete
eigenmodes are stable if the Richardson number, defined as $Ri
\equiv N_{BV}^2/A^2 = N^2/R^2$, is greater then $1/4$.

Equation~(51) describes two important physical phenomena related
to the dynamics of the continuous eigenspectrum, which will be
considered in detail in the two subsequent sections, viz.\ (a)~the
(over)-reflection of the GAW and (b)~the conversion of the EM into
tghe GAW. It will be shown that the EM perturbations can be a very
efficient source of GAW, even for relatively small shearing rates
($Ri > 1/4$).

\section{Over-reflection of Gravito-Alfv\'en waves}

In order to describe the phenomenon of the \emph{over-reflection
}of the GAW in its pure form, let us study Eq.~(51) with ${\cal
I}=0$, yielding:
$$
\frac{{\rm d^2} b}{{\rm d} \tau^2} + \left[1  + \frac{ N^2}{{\cal
K}^2(\tau)} - \frac{R^2}{{\cal K}^4(\tau)}\right] b = 0. \eqno(52)
$$

The standard methods of the analysis of similar equations
\citep{o74} is well-known in quantum mechanics \citep{ll77}. The
behavior of the solution is fully determined by the shear-modified
frequency:
$$
\omega^2(\tau) \equiv 1  + \frac{ N^2}{{\cal K}^2(\tau)} -
\frac{R^2}{{\cal K}^4(\tau)}. \eqno(53)
$$

From the mathematical point of view, the (over)-reflection is
caused by the cavity of $\omega^2(\tau)$ that appears in the
vicinity of the point $\tau^\ast$, where $K_z(\tau^\ast)=0$.

\begin{figure}
\plotone{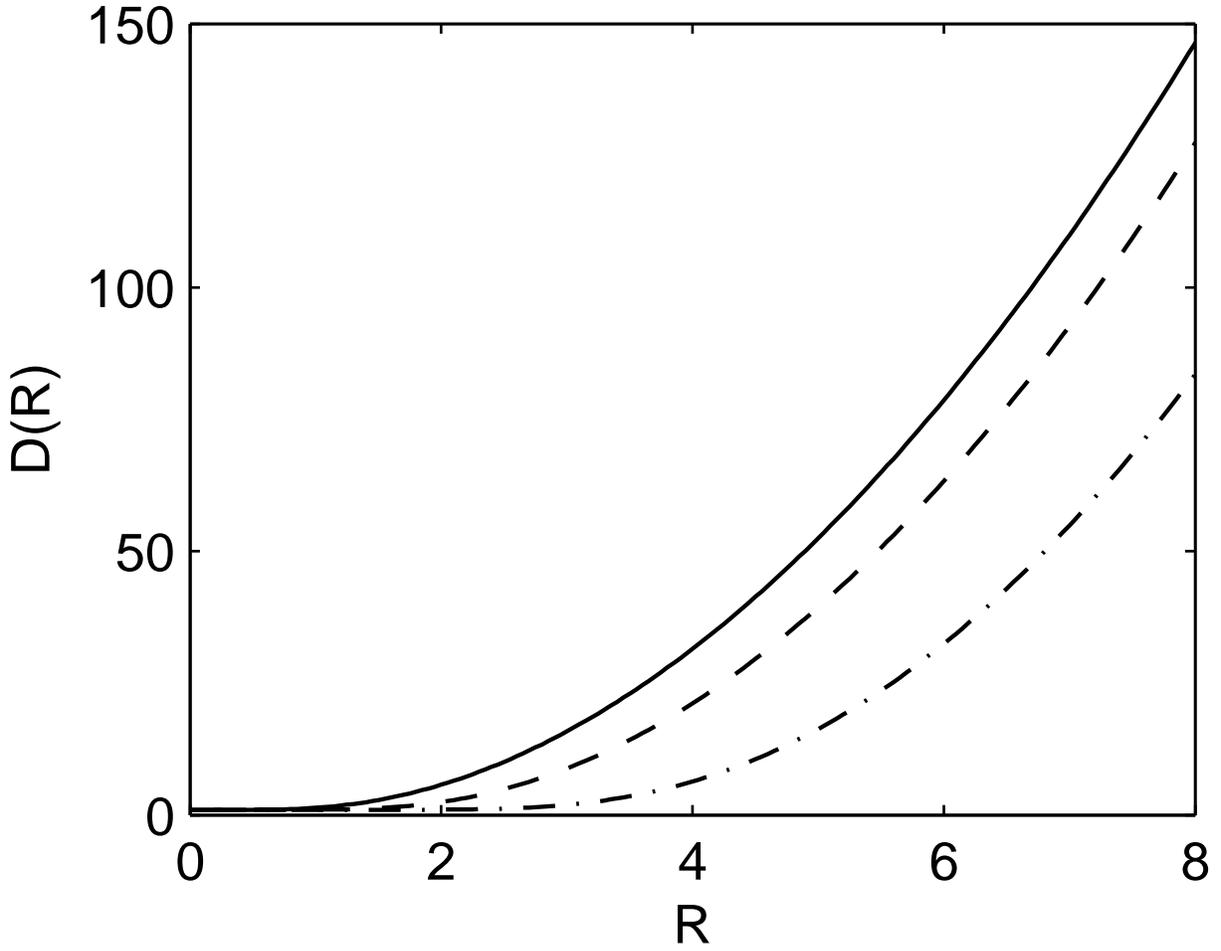} \caption{Dependence of the transmission
coefficient $D(R)$ on the normalized shear parameter $R$ for
different values of the Brunt-V\"ais\"al\"a frequency: $N = 0$
(solid line), $N = 1$ (dashed line) and $N = 2$ (dash-dotted
line).}
\end{figure}

In the areas where $K_z(\tau) \gg R$, the evolution of $b(\tau)$
is adiabatic [${\rm d}\omega(\tau)/{\rm d} \tau \ll
\omega^2(\tau)$] and, therefore, the approximate solution has the
form
$$
b(\tau) \approx  A_+ b_+(\tau) + A_- b_-(\tau), \eqno(54)
$$
where
$$
b_\pm(\tau) = \frac{1}{\sqrt{\omega(\tau)}} e^{\mp i \int
\omega(\tau) d \tau} \eqno(55)
$$
are WKB solutions having  positive and negative phase velocities
along the $x$-axis, respectively, while the $A_\pm$ can be treated
as the intensities of the corresponding types of perturbations.

Let us assume that $A^L_\pm $ and $A^R_\pm $ are the WKB
amplitudes far on the left and right hand side from the point
$\tau^\ast$, respectively. It is well-known \citep{o74} that the
conservation of the Wronskian leads to the following invariant:
$$
|A_+^L|^2 - |A_-^L|^2 = |A_+^R|^2 -
|A_-^R|^2,   \eqno(56)
$$
which physically corresponds to the conservation of the wave
action (see, {e.g.}, \citet{g04} and references therein). If
initially $A_-^L \equiv 0$, the refraction and transmission
coefficients may be defined in the usual way:
$$
Re=\frac{|A_-^R|^2}{|A_+^L|^2},\quad D=\frac{|A_+^R|^2}{|A_+^L|^2},
\eqno(57)
$$
while from Eq.~(56) we obviously have:
$$
1 + Re = D. \eqno(58)
$$

The dependence of the transmission coefficient $D(R)$ on $R$,
obtained through the numerical solution of Eq.~(52) for different
values of the Brunt-V\"ais\"al\"a frequency, viz.\ $N = 0$ (solid
line), $N = 1$ (dashed line) and $N = 2$ (dash-dotted line), is
presented on Fig.~1. The initial conditions for the numerical
solution are chosen from the WKB solutions (55) far on the
left-hand side of the resonant area ($K_z(0) \gg 1,R$). The
analysis leads to the conclusion that the amplification of the
energy density of the perturbations is always finite, but it can
become arbitrarily large with a proper increase of the shearing
rate. It can be seen that the velocity shear can cause a strong
amplification of the GAW, even for moderate values of the
normalized shearing rates.

According to the numerical study of Eq.~(52), the amplitude of the
reflected wave exceeds the amplitude of the incident wave if:
$$
R^2 > 2. \eqno(59)
$$
This phenomenon, originally discovered by Miles for acoustic waves
\citep{m57}, is called \emph{over-reflection}.

Bearing in mind that the energy of the mode is proportional to the
square of its amplitude, we can easily surmise that in the course
of its non-adiabatic evolution the perturbation energy is
increasing. According to Eqs. (56) and (58) the ratio of the total
`post-reflection' energy  to the initial one equals $Re+D>1$. The
energy increases at the expense of the mean (shear) flow energy.
Consequently, over-reflection represents a quite efficient
mechanism for transferring the mean flow energy to perturbations.

In the limit $R \rightarrow 0 $, the reflection coefficient becomes
exponentially small with respect to the parameter $-1/R$, i.e., $Re
\sim \exp(-1/R)$ \citep{g04}.

\section{The generation of Gravito-Alfv\'en waves}

Let us return to the analysis of the full, inhomogeneous equation
(51). When the condition of the adiabatic evolution ${\rm
d}\omega(\tau)/{\rm d} \tau \ll \omega^2(\tau)$ holds, then the
approximate solution of Eq.~(51) has the following form \citep{c94}:
$$
b(\tau) \simeq  A_+ b_+(\tau) + A_- b_-(\tau) + b_E(\tau), \eqno(60)
$$
where the non-periodic solution $b_E(\tau)$ is given by:
$$
b_E(\tau) \simeq \frac{N^2}{{\cal K}(\tau)} \left( 1 +
\frac{N^2}{{\cal K}^2(\tau)} - \frac{R^2}{{\cal K}^4(\tau)}
\right)^{-1}{\cal I}. \eqno(61)
$$
and can be readily identified as the shear-modified \emph{Entropy
Mode} (EM) perturbation.

\begin{figure}
\plotone{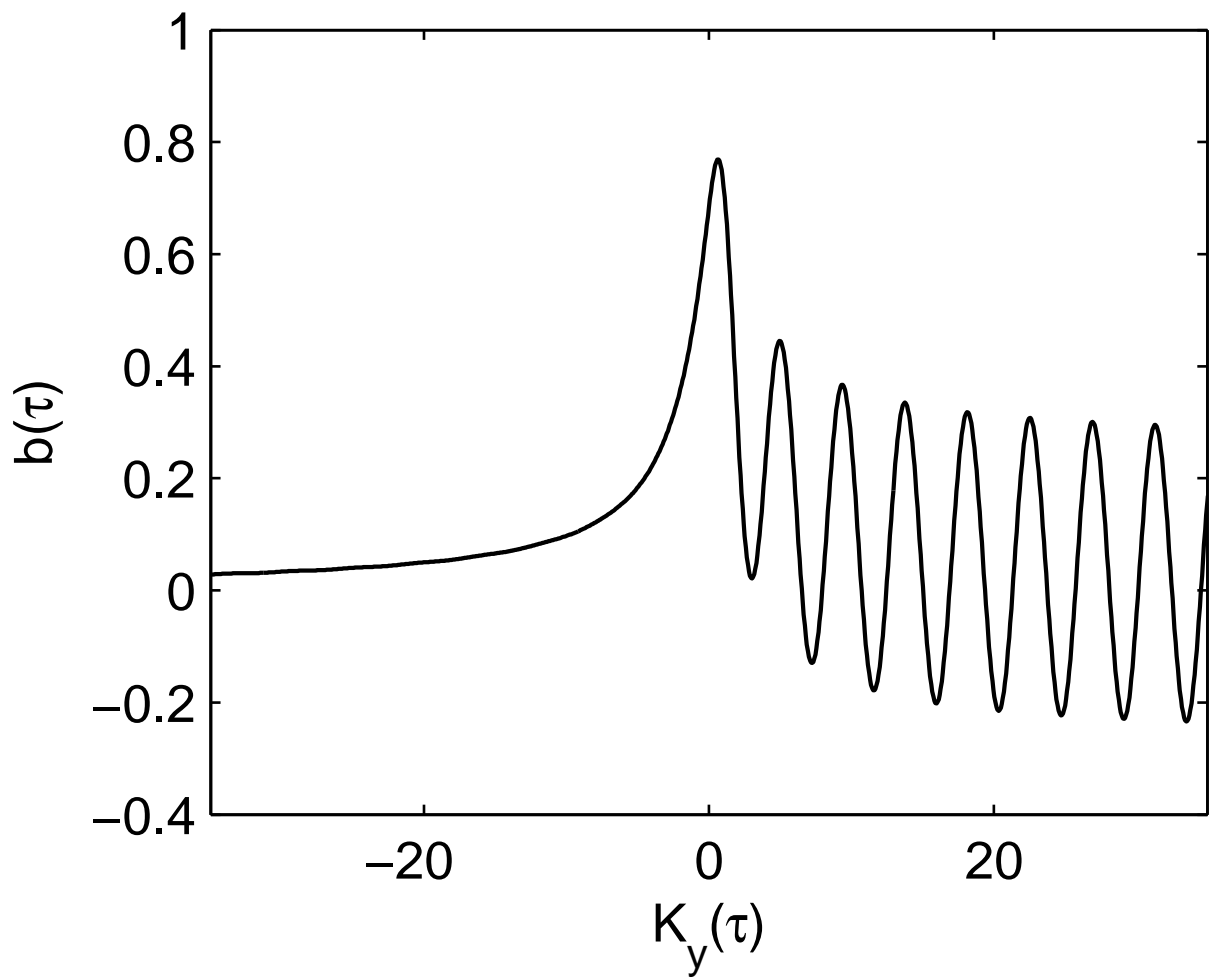} \caption{Temporal evolution of $b(\tau)$ for $N =
1$ and $R=0.7$.}
\end{figure}

It turns out that another physically important phenomenon that
takes place in the system under consideration for moderate and
high shearing rates, is the generation of the GAW by the EM
perturbations. An illustrative example is given by the Fig.~2,
where the numerical solution of Eq.~(51) for the case when $N = 1$
and $R=0.7$ for the variable $b(\tau)$ is shown. The initial
conditions were chosen in such a way that at the beginning (at
$K_z(0) = 35$) there exists only the EM perturbation. Namely, the
initial conditions were found by means of Eqs.~(60) and (61) with
$A_\pm=0$ and ${\cal I}=1$). The figure clearly demonstrates the
generation of the GAW while the EM passes through the interval of
non-adiabatic evolution, situated in the vicinity of the point
${\cal K}_z(\tau^\ast)=0$.

For the quantitative description of the wave generation process,
analogously to the reflection and transmission coefficients
defined in the previous section, we define the \emph{wave
generation coefficient} $G_\pm$ in the following way:
$$
G_\pm =\frac{|A_\pm^L|^2}{|{\cal I}|^2 N^2}. \eqno(62)
$$

With this definition, $G_\pm$ is proportional to the ratio of the
generated wave energy to the energy of incident entropy mode
perturbation. From the symmetry of the problem, it is obvious that
$G_+ = G_- \equiv G$.

In Figure~3, the dependence of the generation coefficient $G(R)$
on the normalized shear parameter $R$ for different values of the
Brunt-V\"ais\"al\"a frequency: $N = 1$ (solid line), $N = 2$
(dashed line) and $N = 3$ (dash-dotted line) is presented.

\begin{figure}
\plotone{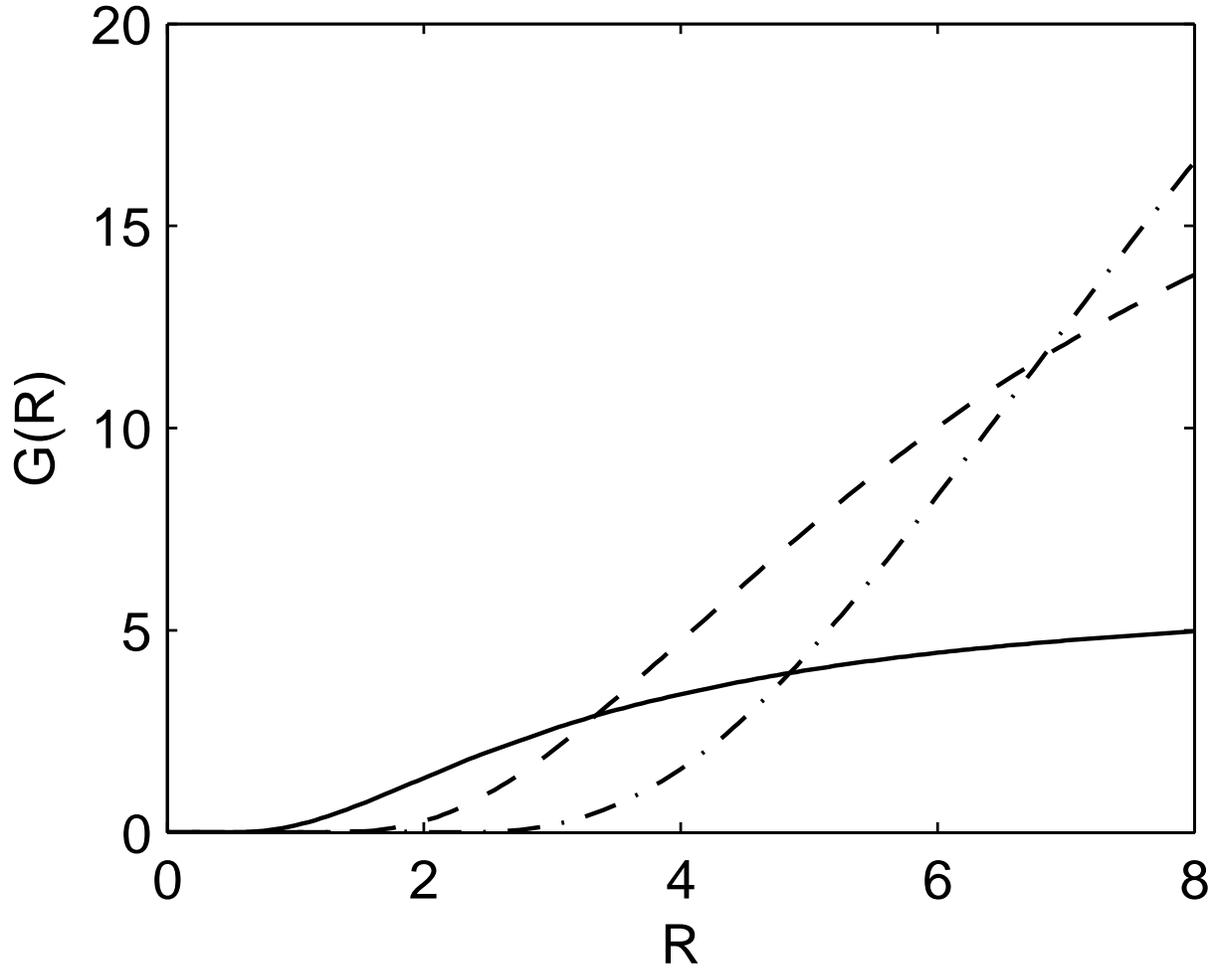} \caption{Dependence of the generation coefficient
$G(R)$ on the normalized shear parameter $R$ for different values
of the Brunt-V\"ais\"al\"a frequency: $N = 1$ (solid line), $N =
2$ (dashed line) and $N = 3$ (dash-dotted line).}
\end{figure}

\begin{figure}
\plotone{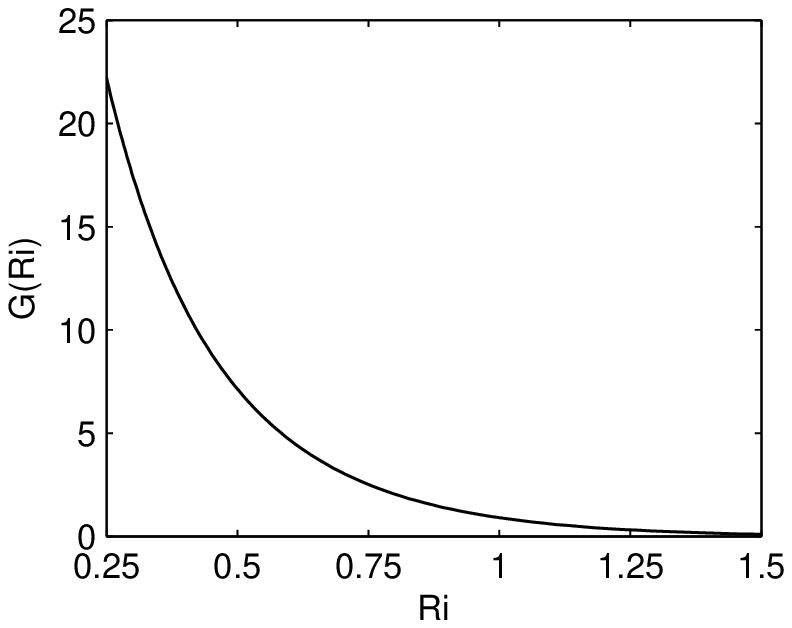} \caption{Dependence of the generation coefficient
$ G(Ri)$ on Richardson number $Ri$ for $N = 4$.}
\end{figure}

\section{Discussion}

In order to explain the near uniformity of the rotation profile in
the radiative region of the Sun and similar regions in solar-type
stars usually two different mechanisms are being proposed. The
first one implies the presence of travelling IGWs generated at the
base of the convective region \citep{ct05}, while the second one
is based on the presumed (and for the Sun observationally
justified) existence of the magnetic field in the radiative zone
\citep{gm98}.

It seems reasonable to assume that the above-mentioned two
mechanisms are not necessarily alternatives, but could be
envisaged as complementary ones. In particular, we may argue that
if magnetic fields are indeed present inside the radiative zone
and the Gravito-MHD waves do pass through the zone, it leads to
intense energy exchange processes between the waves and the
stellar plasma. This circumstance, in its turn, might lead to the
more efficient redistribution of the angular momentum within these
stars, establishing quasi-flat rotation profiles. The presence of
the magnetic field implies that it is more reasonable to speak not
about purely hydrodynamical IGWs, but about Gravito-MHD waves in
general, and Gravito-Alfv\'en waves in particular.

The evolution of the latter mode, present only in the magnetized
case and absent in a purely hydrodynamic medium, is studied in the
present paper. The theoretical background for studying of the
evolution of the full set of Gravito-MHD waves is also developed.

From the general theory of IGWs, we know \citep{dr81} that all the
discrete eigenmodes are stable under the condition that $Ri >
1/4$. Our study shows that the generation of the GAW by the EM
perturbations could provide an efficient mechanism for creating
the required flux of the GAW, even for relatively low shearing
rates, when the discrete eigenmodes are stable. Note, in this
context, that for the case presented in Fig.~2, for instance, the
Richardson number was about $Ri \approx 1.02$.

In the magnetized case, we studied the dependence of the
generation coefficient $G(Ri)$ on the Richardson number $Ri$ for
the relatively weak external magnetic field ($N = 4$). The results
of this study are presented in Fig.~4. The results clearly
indicate that even though the generation coefficient falls rapidly
with increasing $Ri$, the efficient generation of the GAW by the
EM perturbations still takes place for $Ri < 1.25$.

Obviously, our study has a simplified nature and it only indicates
at the possibly important role of \emph{non-modal} phenomena
related with GAW in the redistribution of the angular momentum and
the flattening of the rotation curves in solar-type stars. For
making this claim more convincing, we have to consider the
compressible, three-dimensional analogue of this problem and we
have to verify whether the full set of Gravito-MHD waves is as
efficient (or even more efficient) in exchanging energy with the
equilibrium flow. We know from previous studies that in the
absence of the gravity-induced stratification among the
shear-modified MHD waves, the fast magnetosonic waves are the most
efficient ones in exchanging energy with the equilibrium flow
\citep{prm99,rpm00}. It is of great interest to check whether the
fast Gravito-magnetosonic waves have the same quality and to
determine what their contribution in the angular momentum
redistribution could be. The study of the linear dynamics of all
Gravito-MHD waves is currently initiated and the results will be
published in a subsequent paper.

Finally, one has to remember that the non-modal approach that has
been applied in the present paper, provides no information about
the spatial aspects of the shear-induced processes, because the
study of the Spatial Fourier Harmonics is confined to the phase
space of the wave number vectors ${\bf k}(t)$. In order to have a
clear idea about the spatial appearance of these processes, one
has to study them numerically (similarly to the uniform,
non-stratified flow study made by \citet{b01}) and to check how
the non-modal phenomena couple with the traditional rotational
instabilities.

\acknowledgments
These results were obtained in the framework of the projects GOA
2004/01 (K.U.Leuven), G.0304.07 (FWO-Vlaanderen) and C90203 (ESA
Prodex 8). Andria Rogava wishes to thank the \emph{Katholieke
Universiteit Leuven} (Leuven, Belgium) and the \emph{Abdus Salam
International Centre for Theoretical Physics}(Trieste, Italy) for
supporting him, in part, through a Senior Postdoctoral Fellowship
and Senior Associate Membership Award, respectively. The research
of Andria Rogava and Grigol Gogoberidze was supported in part by
the Georgian National Science Foundation grant GNSF/ST06/4-096.
The research of Grigol Gogoberidze was supported in part by the
INTAS grant 06-1000017-9258.



\end{document}